\documentclass{article}

\usepackage{amsmath}
\usepackage{arxiv}
\usepackage{natbib}
\usepackage[utf8]{inputenc} 
\usepackage[T1]{fontenc}    
\usepackage{hyperref}       
\usepackage{url}            
\usepackage{booktabs}       
\usepackage{amsfonts}       
\usepackage{nicefrac}       
\usepackage{microtype}      
\usepackage{lipsum}
\usepackage{graphicx}
\graphicspath{ {./images/} }

\title{Deep Learning-Based Monthly Temperature Prediction for Jilin Province: A Multi-Model Comparative Study 2000–2026}

\author{
 Xingyue Deng \thanks{Corresponding author} \\
  College of Geography and Ocean Sciences\\
  Yanbian University\\
  Hunchun, Jilin 133300 \\
  \texttt{34436832632@qq.com} \\
 \And
 Xuechen Liang \\
}

\begin{document}
\maketitle
\begin{abstract}
Jilin Province, a core commercial grain production base in China, is characterized by a mid-temperate continental monsoon climate with significant temperature fluctuations, where agricultural production and ecological security are highly dependent on temperature conditions. Existing temperature prediction studies are mostly concentrated on national or southeastern coastal regions, with few targeting the specific climatic characteristics of Jilin Province, and most models fail to fully integrate the spatiotemporal differentiation and seasonal periodicity of local temperature, leading to limited prediction accuracy.

Based on the 1 km × 1 km monthly mean temperature raster data of Jilin Province from 2000 to 2024, this study first systematically analyzed the spatiotemporal variation characteristics of regional temperature, then constructed a multi-model comparison system including four deep learning models (LSTM, GRU, BiLSTM, Transformer) and five traditional machine learning models (Ridge Regression, Lasso Regression, SVR, Random Forest, Gradient Boosting). The model performance was evaluated using Root Mean Squared Error (RMSE), Mean Absolute Error (MAE), and Coefficient of Determination ($R^2$).

The results show that the temperature in Jilin Province presents an obvious latitudinal zonal distribution and a significant warming trend, with strong seasonal periodicity and high temporal autocorrelation. Among all models, the Long Short-Term Memory (LSTM) model achieves the optimal prediction performance, with test set RMSE = 2.26 $^\circ$C, MAE = 1.83 $^\circ$C, and $R^2$ = 0.9655, significantly outperforming traditional machine learning models and the Transformer model. Using the optimized LSTM model, the temperature of Jilin Province in 2025–2026 was predicted, revealing that the regional temperature will maintain stable seasonal fluctuations with an annual mean temperature of approximately 4.9 $^\circ$C.

This study enriches the temperature prediction research system for cold regions in mid-latitudes, verifies the applicability of the LSTM model in monthly temperature prediction for Jilin Province, and provides scientific support for local agricultural production planning, frost disaster early warning, and extreme temperature risk prevention.

\end{abstract}


\section{Introduction}

As an important commercial grain production base in China, agricultural production in Jilin Province is highly dependent on climatic conditions, particularly the seasonal variation of temperature, which directly determines the entire cycle of crop sowing, growth, and maturation. Accurate temperature prediction can provide scientific support for crop variety selection, planting structure adjustment, and frost disaster early warning, thereby reducing agricultural losses caused by meteorological disasters \cite{yu2025dual}. Furthermore, long-term temperature trends and extreme temperature events in Jilin Province affect regional ecosystem stability, water resource allocation, and energy supply-demand balance \cite{zhang2025analysis}. Therefore, conducting time series prediction research on temperature in Jilin Province holds both important theoretical value and practical significance \cite{masson2021climate}.

In recent years, with the rapid development of meteorological observation technology, remote sensing inversion technology, and data processing technology, obtaining high-precision, long-term temperature data has become feasible, providing solid data support for temperature prediction research \cite{mohamed2025hybrid}. Simultaneously, the rise of machine learning and deep learning technologies has provided new methodological approaches for capturing nonlinear characteristics and modeling long-term dependencies in time series data \cite{wang2021deep, wang2019application}, breaking through the limitations of traditional statistical prediction methods in complex temperature series prediction and continuously improving prediction accuracy \cite{nevavuori2020crop, he2025deep}. However, existing research mostly focuses on national scales or southeastern coastal regions, with specialized temperature prediction studies targeting the cold and continental monsoon climate characteristics of Jilin Province being relatively scarce \cite{mutinda2025forecasting}. Moreover, most studies have not fully integrated the spatiotemporal differentiation characteristics and seasonal periodicity of temperature in Jilin Province, resulting in lack of specificity in model selection and parameter settings, leading to prediction accuracy that fails to meet practical application requirements \cite{zhang2025analysis}.

Accurate temperature prediction can provide scientific guidance for agricultural production in Jilin Province, helping agricultural departments rationally plan planting layouts, select suitable crop varieties, and provide early warnings for frost and extreme high temperature disasters, thereby reducing economic losses caused by disasters and ensuring regional food security \cite{yu2025dual, mohamed2025hybrid}. Second, temperature prediction results can provide decision-making basis for regional ecological environment protection and water resource management, assisting Jilin Province in addressing climate change and optimizing ecological restoration plans to maintain ecosystem stability \cite{masson2021climate, zhang2025analysis}. Additionally, temperature prediction can serve livelihood areas such as energy dispatching and urban heating/cooling planning, optimizing energy allocation, reducing energy consumption, and improving livelihood security levels \cite{zhang2021elevation}. Furthermore, temperature prediction results for 2025-2026 can provide reference for Jilin Province in responding to extreme temperature events, enabling early formulation of prevention and control measures to reduce the impact of extreme temperatures on residents' lives, transportation, and public safety \cite{xue2021change}.

Under the background of global climate change, research on regional temperature changes has become a hot topic in meteorology and environmental science. Domestic and international scholars have conducted extensive research on temperature change characteristics across different regions and time scales, confirming the significant upward trend of global temperature with obvious regional differences \cite{masson2021climate}. In Northeast China, existing research indicates that the average annual temperature increase rate from 1961 to 2017 was 0.31$^{\circ}$C/10 years, higher than the national average during the same period, with winter warming rates significantly higher than summer, showing a ``warm winter'' characteristic \cite{zhang2021elevation, chun2013changes, xue2021change}. Focusing on Jilin Province, some scholars have concentrated on interannual and seasonal temperature variation characteristics, analyzing the influence of topography and atmospheric circulation on temperature distribution. They found that temperature in Jilin Province shows a significant upward trend, with spatial distribution presenting latitudinal zonal characteristics decreasing from south to north, significantly influenced by topographic differentiation \cite{zhang2025analysis}. However, existing research mostly focuses on temperature change analysis before 2020, with relatively scarce research on the spatiotemporal differentiation characteristics of long-term temperature series from 2000 to 2024. Moreover, the analysis of driving factors behind temperature change trends is insufficient, making it difficult to comprehensively reflect recent temperature change patterns in Jilin Province \cite{zhang2025analysis, zhang2021elevation}.ly reflect recent temperature change patterns in Jilin Province .

Temperature prediction is essentially a time series prediction problem, and its research methods have mainly evolved through three stages: traditional statistical prediction, machine learning prediction, and deep learning prediction. Traditional temperature prediction methods are dominated by statistical models, including linear regression, ARIMA models, and exponential smoothing methods. These methods are simple in principle and computationally efficient but struggle to capture nonlinear characteristics and long-term dependency relationships in temperature series, resulting in limited prediction accuracy, particularly unsuitable for temperature prediction in complex climate zones. With the development of machine learning technology, traditional machine learning models such as Support Vector Regression (SVR), Random Forest, and Gradient Boosting Trees have been widely applied in temperature prediction. By mining nonlinear relationships between temperature series and influencing factors, these models have improved prediction accuracy. However, they still have limitations in handling long-term series and effectively capturing seasonal periodicity of temperature .

However, existing temperature prediction research still has several shortcomings: First, most studies focus on daily-scale or hourly-scale temperature prediction, with relatively few studies on monthly-scale temperature prediction, while monthly-scale temperature prediction holds greater practical value for agricultural planning and seasonal disaster early warning . Second, model selection lacks specificity; most studies do not select appropriate models based on regional climate characteristics, resulting in failure to fully exploit model advantages. For example, Transformer models are prone to overfitting with small sample sizes, making them unsuitable for medium-small scale temperature datasets. Third, specialized research on cold and continental monsoon climate zones is limited, with existing model parameter settings mostly referencing other regions without fully considering the characteristics of large temperature fluctuations and strong periodicity in such zones, resulting in prediction accuracy that fails to meet practical requirements . Fourth, multi-model comparison research is not comprehensive enough; most studies only compare a few models without systematically evaluating performance differences between deep learning and traditional machine learning models, making it difficult to identify optimal prediction models.

Research Innovations
\begin{enumerate}
    \item \textbf{Innovation in Research Perspective}: Focusing on the mid-temperate continental monsoon climate zone of Jilin Province, this study systematically analyzes the spatiotemporal differentiation characteristics and variation trends of temperature based on long-term monthly-scale temperature data from 2000 to 2024. This compensates for the deficiency of existing research on long-term temperature series in this region and demonstrates strong specificity.
    
    \item \textbf{Innovation in Model System}: This study constructs a multi-model comparison framework covering both deep learning and traditional machine learning, comprehensively evaluating the performance of nine typical prediction models. It clarifies the applicability of different models in temperature prediction in Jilin Province, avoiding the limitations of single-model prediction and providing reference for temperature prediction model selection in similar regions.
    
    \item \textbf{Innovation in Methodological Optimization}: Combining the periodicity and strong autocorrelation characteristics of temperature data in Jilin Province, this research optimizes data preprocessing procedures and model training strategies. It employs sliding window reconstruction to capture seasonal periodicity and uses regularization and early stopping strategies to prevent model overfitting, improving the accuracy and stability of temperature prediction.
    
    \item \textbf{Innovation in Application Value}: Based on the optimal model, this study conducts temperature prediction for 2025-2026 and proposes targeted suggestions for extreme temperature risk prevention. It provides precise decision-making support for agricultural production, ecological protection, and livelihood security in Jilin Province, demonstrating outstanding practical value.
\end{enumerate}

\section{Study Area and Data}
\subsection{Overview of the Study Area}
Jilin Province is located in the central part of Northeast China (40$^\circ$52'—46$^\circ$18' N, 121$^\circ$38'—131$^\circ$19' E), with a total area of 187,400 square kilometers. It lies in the mid-latitude temperate continental monsoon climate zone. The regional climate is characterized by long, cold winters, with average January temperatures as low as -18$^\circ$C to -10$^\circ$C, and warm, short summers, with average July temperatures of 20$^\circ$C to 23$^\circ$C. The drastic temperature fluctuations in spring and autumn, with an annual temperature range exceeding 40$^\circ$C, make Jilin Province an ideal testbed for validating time series forecasting models. Furthermore, as a major commodity grain base in China, accurate temperature prediction is of significant practical value for agricultural planning, crop variety selection, and frost disaster warnings.

\subsection{Data Sources}
The temperature data used in this study is a monthly mean air temperature raster dataset for Jilin Province from January 2000 to December 2024, with a spatial resolution of $1\text{km}\times1\text{km}$. The data originates from meteorological reanalysis products, spatially interpolated by the National Meteorological Science Data Center. The dataset is generated by integrating ground meteorological station observations with Digital Elevation Models (DEM) and remote sensing inversion data, using the Thin Plate Spline interpolation method, ensuring high accuracy and spatiotemporal continuity.

\subsection{Data Preprocessing}
\subsubsection{Study Area Clipping}
Using the administrative boundary vector data of Jilin Province (1:250,000 scale from the National Geomatics Center of China) as a mask, the study area was extracted from the national temperature raster data:
\begin{equation}
T_{\text{Jilin}} = T_{\text{national}} \cap \text{Boundary}_{\text{Jilin}}
\end{equation}
This operation ensures that subsequent analyses are strictly confined within the administrative boundaries of Jilin Province, eliminating interference from surrounding regions.

\subsubsection{Spatial Sampling Strategy}
Considering the massive volume of the full raster data (approximately 18,000 pixels) and the high spatial autocorrelation of adjacent pixels, a systematic sampling method was employed to uniformly deploy 30 sampling points across Jilin Province. The sampling strategy adhered to the following principles: (1) Uniform spatial distribution covering the eastern, central, and western climatic sub-regions; (2) Avoidance of large water bodies and urban heat island areas; (3) Representativeness of elevation gradients, encompassing plain (<200 m), hilly (200—500 m), and mountainous (>500 m) landforms. Monthly temperature time series were extracted for each sampling point, constructing a dataset of 4,680 samples (30 points $\times$ 156 months).

\subsubsection{Data Partitioning}
The dataset was strictly partitioned into training and testing sets in chronological order to ensure the evaluation aligns with the real-world prediction scenario (i.e., predicting the future from the past).
\begin{itemize}
    \item \textbf{Training Set:} January 2000 – December 2019 (20 years, 3,600 samples, 76.9\%).
    \item \textbf{Testing Set:} January 2020 – December 2024 (5 years, 1,080 samples, 23.1\%).
\end{itemize}
This ratio balances the adequacy of model training with the reliability of independent validation.

\subsubsection{Normalization}
Min-Max scaling was applied to map the temperature data to the $[0,1]$ interval:
\begin{equation}
x_{\text{norm}} = \frac{x - x_{\min}}{x_{\max} - x_{\min}}
\end{equation}
Here, $x_{\min}$ and $x_{\max}$ were determined based on the training set statistics. The same scaling parameters were applied to the testing set to prevent data leakage. Normalization accelerates neural network convergence and eliminates dimensional effects.

\subsubsection{Sequence Reconstruction}
Temperature prediction is fundamentally a time series forecasting problem. A sliding window approach was used to restructure the raw sequence into a supervised learning format. An input window length of $L=12$ months (utilizing the past year's records) was set to predict the next month:
\begin{equation}
X_t = \{x_{t-L+1}, x_{t-L+2}, ..., x_t\} \in \mathbb{R}^L
\end{equation}
\begin{equation}
y_t = x_{t+1} \in \mathbb{R}
\end{equation}
With a sliding step of 1, this generated 4,668 effective samples (144 months $\times$ 30 points). This setting ensures sufficient sample volume while enabling the model to capture the seasonal periodicity of temperature.

\subsubsection{Data Quality Analysis}
Exploratory analysis of the training set revealed a monthly mean temperature range of -22.3$^\circ$C to 26.8$^\circ$C, an annual mean of 3.8$^\circ$C, a standard deviation of 12.4$^\circ$C, and a skewness coefficient of -0.32, indicating a slight left-skewed distribution. Autocorrelation analysis showed a coefficient of 0.87 at a 12-month lag, indicating a significant annual cycle, and 0.92 at a 1-month lag, confirming high temporal correlation. These statistical features provide crucial guidance for model design: the strong autocorrelation supports the use of Recurrent Neural Networks (RNNs) to capture temporal dependencies, while the significant periodicity suggests the model requires long-term memory capabilities.

\subsection{Model Architecture}
This study establishes a temperature forecasting model system comprising four deep learning models: Long Short-Term Memory (LSTM), Gated Recurrent Unit (GRU), Bidirectional LSTM (BiLSTM), and Transformer. Additionally, five traditional machine learning models—Ridge Regression, Lasso Regression, Support Vector Regression (SVR), Random Forest, and Gradient Boosting—are included as baseline benchmarks.

\subsubsection{Long Short-Term Memory (LSTM)}
The LSTM is a specialized Recurrent Neural Network (RNN) that mitigates the vanishing gradient problem through gating mechanisms, making it suitable for capturing long-term dependencies in time series.

\subsubsubsection{LSTM Unit Structure}
The core components are the Cell State and three gates: Forget Gate, Input Gate, and Output Gate.
The Forget Gate controls the retention of historical information:
\begin{equation}
f_t = \sigma(W_f \cdot [h_{t-1}, x_t] + b_f)
\end{equation}
where $\sigma(\cdot)$ is the Sigmoid activation function, $W_f$ is the weight matrix, $b_f$ is the bias, $h_{t-1}$ is the previous hidden state, and $x_t$ is the current input.
The Input Gate determines the storage ratio of new information:
\begin{equation}
i_t = \sigma(W_i \cdot [h_{t-1}, x_t] + b_i)
\end{equation}
The candidate cell state is generated via the hyperbolic tangent function:
\begin{equation}
\tilde{C}_t = \tanh(W_C \cdot [h_{t-1}, x_t] + b_C)
\end{equation}
The cell state is updated by combining historical and current information:
\begin{equation}
C_t = f_t \odot C_{t-1} + i_t \odot \tilde{C}_t
\end{equation}
where $\odot$ denotes the Hadamard product (element-wise multiplication).
The Output Gate controls the hidden state output:
\begin{equation}
o_t = \sigma(W_o \cdot [h_{t-1}, x_t] + b_o)
\end{equation}
\begin{equation}
h_t = o_t \odot \tanh(C_t)
\end{equation}

\begin{figure}[!ht]  
    \centering  
    \includegraphics[width=0.8\textwidth]{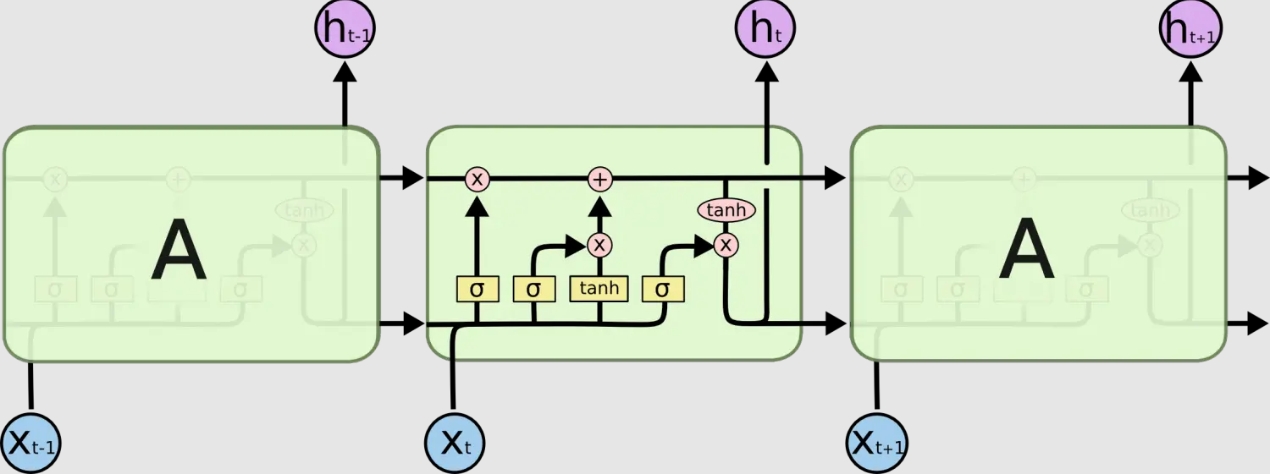}  
    \caption{LSTM architecture}  
    \label{fig:my-figure}  
\end{figure}

\subsubsubsection{Network Configuration}
The LSTM architecture consists of two stacked LSTM layers, each with 64 hidden units. A Dropout rate of 0.1 was applied between layers to prevent overfitting. The input sequence length is 12, and the output dimension is 1, corresponding to the prediction for the next month.

\subsubsection{Gated Recurrent Unit (GRU)}
The GRU is a simplified variant of LSTM that merges the forget and input gates into a single Update Gate, reducing parameters and improving computational efficiency.

\subsubsubsection{GRU Mathematical Formulation}
The GRU contains a Reset Gate and an Update Gate:
\begin{equation}
r_t = \sigma(W_r \cdot [h_{t-1}, x_t])
\end{equation}
\begin{equation}
z_t = \sigma(W_z \cdot [h_{t-1}, x_t])
\end{equation}
The candidate hidden state is influenced by the Reset Gate:
\begin{equation}
\tilde{h}_t = \tanh(W \cdot [r_t \odot h_{t-1}, x_t])
\end{equation}
The final hidden state is a weighted combination of the previous and candidate states:
\begin{equation}
h_t = (1 - z_t) \odot h_{t-1} + z_t \odot \tilde{h}_t
\end{equation}


\subsubsubsection{Network Configuration}
The GRU network adopts the same layer count and hidden unit configuration (2 layers, 64 units) as the LSTM for a fair performance comparison.

\begin{figure}[!ht]  
    \centering  
    \includegraphics[width=0.8\textwidth]{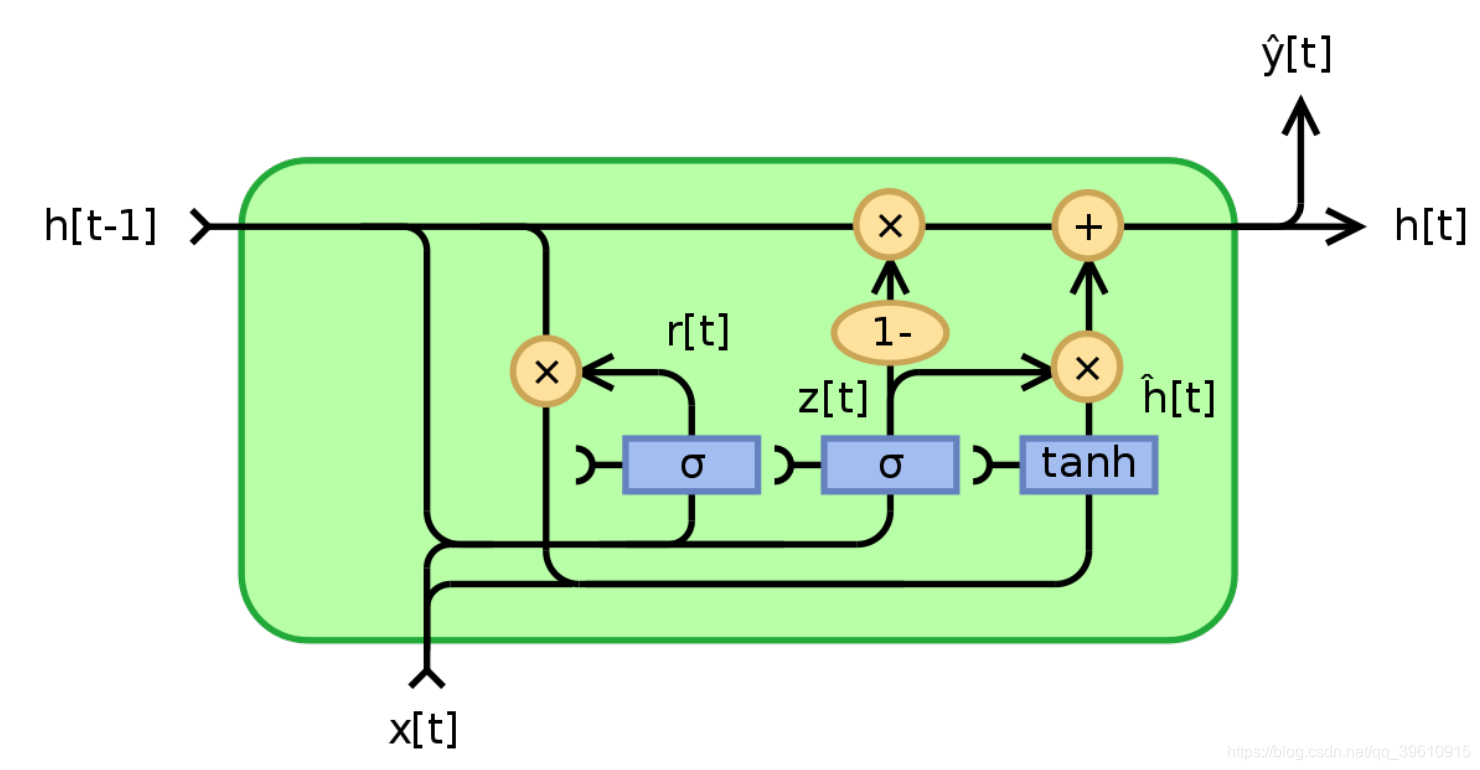}  
    \caption{GRU architecture}  
    \label{fig:my-figure}  
\end{figure}

\subsubsection{Bidirectional Long Short-Term Memory (BiLSTM)}
BiLSTM models temporal dependencies in both forward and backward directions, leveraging contextual information from the past and future.

\subsubsubsection{Bidirectional Mechanism}
The forward LSTM processes the sequence from past to future:
\begin{equation}
\overrightarrow{h}_t = \text{LSTM}(x_t, \overrightarrow{h}_{t-1})
\end{equation}
The backward LSTM processes the sequence from future to past:
\begin{equation}
\overleftarrow{h}_t = \text{LSTM}(x_t, \overleftarrow{h}_{t+1})
\end{equation}
The final hidden state is obtained by concatenating the bidirectional outputs:
\begin{equation}
h_t = [\overrightarrow{h}_t; \overleftarrow{h}_t]
\end{equation}
where $[;]$ denotes vector concatenation.

\begin{figure}[!ht]  
    \centering  
    \includegraphics[width=0.8\textwidth]{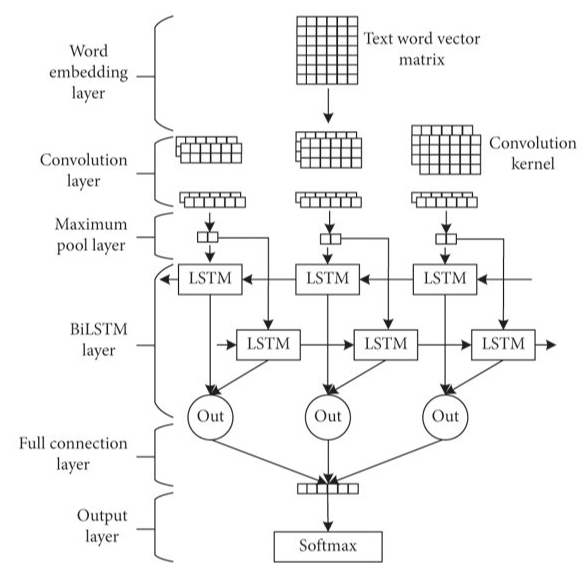}  
    \caption{BiLSTM architecture}  
    \label{fig:my-figure}  
\end{figure}

\subsubsubsection{Network Configuration}
The BiLSTM network contains 64 forward and 64 backward units per layer, resulting in a total hidden dimension of 128. While the training utilizes bidirectional context to maximize data utility, inference strictly relies on historical information, as future data is unavailable in real-world forecasting.

\subsubsection{Transformer Model}
The Transformer relies entirely on attention mechanisms, discarding recurrence, to capture global dependencies within sequences.

\subsubsubsection{Self-Attention Mechanism}
Scaled dot-product attention computes the correlation between Query ($Q$), Key ($K$), and Value ($V$):
\begin{equation}
\text{Attention}(Q,K,V) = \text{softmax}\left( \frac{QK^T}{\sqrt{d_k}} \right)V
\end{equation}
where $d_k$ is the key dimension. The scaling factor $\sqrt{d_k}$ prevents softmax gradient vanishing.
Multi-head attention enhances representational power via parallel attention computations:
\begin{equation}
\text{MultiHead}(Q,K,V) = \text{Concat}(\text{head}_1, ..., \text{head}_h)W^O
\end{equation}
\begin{equation}
\text{head}_i = \text{Attention}(QW_i^Q, KW_i^K, VW_i^V)
\end{equation}

\begin{figure}[!ht]  
    \centering  
    \includegraphics[width=0.8\textwidth]{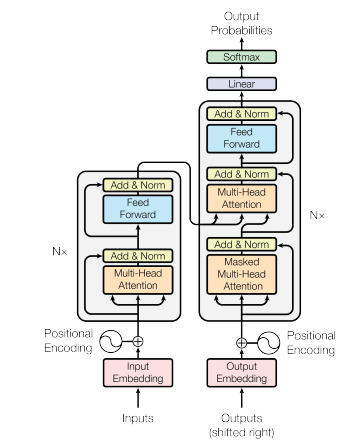}  
    \caption{Transformer architecture}  
    \label{fig:my-figure}  
\end{figure}

\subsubsubsection{Feed-Forward Network and Normalization}
Each encoder layer includes a fully connected feed-forward network:
\begin{equation}
\text{FFN}(x) = \max(0, xW_1 + b_1)W_2 + b_2
\end{equation}
Layer Normalization stabilizes the training process:
\begin{equation}
\text{LayerNorm}(x) = \gamma \cdot \frac{x - \mu}{\sigma} + \beta
\end{equation}

\subsubsubsection{Positional Encoding}
Since the Transformer lacks inherent sequence order perception, sinusoidal positional encoding is introduced:
\begin{equation}
PE_{(pos,2i)} = \sin\left( \frac{pos}{10000^{2i/d_{\text{model}}}} \right)
\end{equation}
\begin{equation}
PE_{(pos,2i+1)} = \cos\left( \frac{pos}{10000^{2i/d_{\text{model}}}} \right)
\end{equation}

\subsubsubsection{Network Configuration}
The Transformer encoder comprises 2 layers, with a model dimension of 64, a feed-forward dimension of 256, 4 attention heads, and a Dropout rate of 0.1.

\subsubsection{Traditional Machine Learning Models}
To comprehensively evaluate the performance advantages of deep learning models, five representative traditional machine learning models were selected as baselines:

\subsubsubsection{Ridge Regression}
Ridge regression introduces L2 regularization to the ordinary least squares loss:
\begin{equation}
\hat{\beta}_{\text{ridge}} = \arg\min_{\beta} \left\{ \sum_{i=1}^n (y_i - \beta_0 - \sum_{j=1}^p x_{ij}\beta_j)^2 + \lambda \sum_{j=1}^p \beta_j^2 \right\}
\end{equation}
where $\lambda$ is the regularization parameter determined via cross-validation.

\subsubsubsection{Lasso Regression}
Lasso regression employs L1 regularization, which performs feature selection:
\begin{equation}
\hat{\beta}_{\text{lasso}} = \arg\min_{\beta} \left\{ \sum_{i=1}^n (y_i - \beta_0 - \sum_{j=1}^p x_{ij}\beta_j)^2 + \lambda \sum_{j=1}^p |\beta_j| \right\}
\end{equation}

\subsubsubsection{Support Vector Regression (SVR)}
SVR performs nonlinear regression using an $\epsilon$-insensitive loss function and a kernel. The Radial Basis Function (RBF) kernel was used:
\begin{equation}
K(x,x') = \exp(-\gamma \|x - x'\|^2)
\end{equation}

\subsubsubsection{Random Forest}
Random Forest integrates multiple decision trees via Bagging:
\begin{equation}
\hat{y} = \frac{1}{T} \sum_{t=1}^T h_t(x)
\end{equation}
where $T$ is the number of trees and $h_t(x)$ is the prediction of the $t$-th tree.

\subsubsubsection{Gradient Boosting}
Gradient Boosting trains weak learners iteratively to fit the residuals of the previous model:
\begin{equation}
F_m(x) = F_{m-1}(x) + \eta \cdot h_m(x)
\end{equation}
where $\eta$ is the learning rate, controlling the contribution of each tree.

\subsection{Model Training Strategy}
\subsubsection{Loss Function}
Mean Squared Error (MSE) was adopted as the optimization objective:
\begin{equation}
L_{\text{MSE}} = \frac{1}{N} \sum_{i=1}^N (y_i - \hat{y}_i)^2
\end{equation}
MSE penalizes larger errors more heavily, aligning with the requirement for accuracy in predicting extreme temperature events.

\subsubsection{Optimization Algorithm}
Deep learning models were optimized using the Adam algorithm, which combines the advantages of Momentum and RMSProp to adaptively adjust parameter learning rates:
\begin{equation}
m_t = \beta_1 m_{t-1} + (1 - \beta_1)g_t
\end{equation}
\begin{equation}
v_t = \beta_2 v_{t-1} + (1 - \beta_2)g_t^2
\end{equation}
\begin{equation}
\hat{m}_t = \frac{m_t}{1 - \beta_1^t}, \quad \hat{v}_t = \frac{v_t}{1 - \beta_2^t}
\end{equation}
\begin{equation}
\theta_t = \theta_{t-1} - \frac{\eta}{\sqrt{\hat{v}_t} + \epsilon} \hat{m}_t
\end{equation}
where $g_t$ is the gradient, $\beta_1 = 0.9$, $\beta_2 = 0.999$, the initial learning rate $\eta = 0.005$, and $\epsilon = 10^{-8}$ is a smoothing term to prevent division by zero.

\subsubsection{Learning Rate Scheduling}
A ReduceLROnPlateau strategy was employed to dynamically adjust the learning rate: if the validation loss did not improve for 10 consecutive epochs (patience), the learning rate was multiplied by a decay factor of 0.5:
\begin{equation}
\eta_{\text{new}} = \eta_{\text{old}} \times 0.5
\end{equation}
This strategy aids in fine-tuning parameters during the later stages of training and prevents oscillation.

\subsubsection{Regularization and Early Stopping}
To prevent overfitting, the following measures were taken: (1) Dropout: A rate of 0.1 was applied between LSTM, GRU, and BiLSTM layers; (2) Early Stopping: Training was terminated if the validation loss did not decrease for 20 consecutive epochs, and the model parameters with the best validation performance were restored.

\subsection{Evaluation Metrics}
To comprehensively assess the predictive performance of the models, the following three metrics were adopted:

\subsubsubsection{Root Mean Squared Error (RMSE)}
\begin{equation}
\text{RMSE} = \sqrt{\frac{1}{N} \sum_{i=1}^N (y_i - \hat{y}_i)^2}
\end{equation}
RMSE is sensitive to outliers and reflects the average deviation between predicted and observed values in degrees Celsius ($^\circ$C).

\subsubsubsection{Mean Absolute Error (MAE)}
\begin{equation}
\text{MAE} = \frac{1}{N} \sum_{i=1}^N |y_i - \hat{y}_i|
\end{equation}
MAE intuitively reflects the average magnitude of prediction errors and is more robust to outliers than RMSE.

\subsubsubsection{Coefficient of Determination ($R^2$)}
\begin{equation}
R^2 = 1 - \frac{\sum_{i=1}^N (y_i - \hat{y}_i)^2}{\sum_{i=1}^N (y_i - \bar{y})^2}
\end{equation}
where $\bar{y}$ is the mean of the observed values. $R^2 \in (-\infty, 1]$, with values closer to 1 indicating stronger explanatory power. A negative value indicates the model performs worse than a simple mean predictor.

\section{Experimental Results and Analysis}

\subsection{Data Characteristics Analysis}

\subsubsection{Temperature Statistical Features}

This study is based on monthly mean temperature data from 30 uniformly distributed sampling points in Jilin Province from January 2000 to December 2024, totaling 4,680 monthly temperature records. Table~\ref{tab:temp_statistics} presents the basic statistical characteristics of the temperature data.

\begin{table}[htbp]
\centering
\caption{Statistical characteristics of temperature data in Jilin Province}
\label{tab:temp_statistics}
\begin{tabular}{lc}
\toprule
\textbf{Statistical Indicator} & \textbf{Value} \\
\midrule
Mean Temperature & $-4.62^\circ\text{C}$ \\
Standard Deviation & $14.71^\circ\text{C}$ \\
Maximum Temperature & $26.00^\circ\text{C}$ \\
Minimum Temperature & $-21.50^\circ\text{C}$ \\
Temperature Range & $47.50^\circ\text{C}$ \\
\bottomrule
\end{tabular}
\end{table}

\begin{figure}[htbp]
\centering
\includegraphics[width=\textwidth]{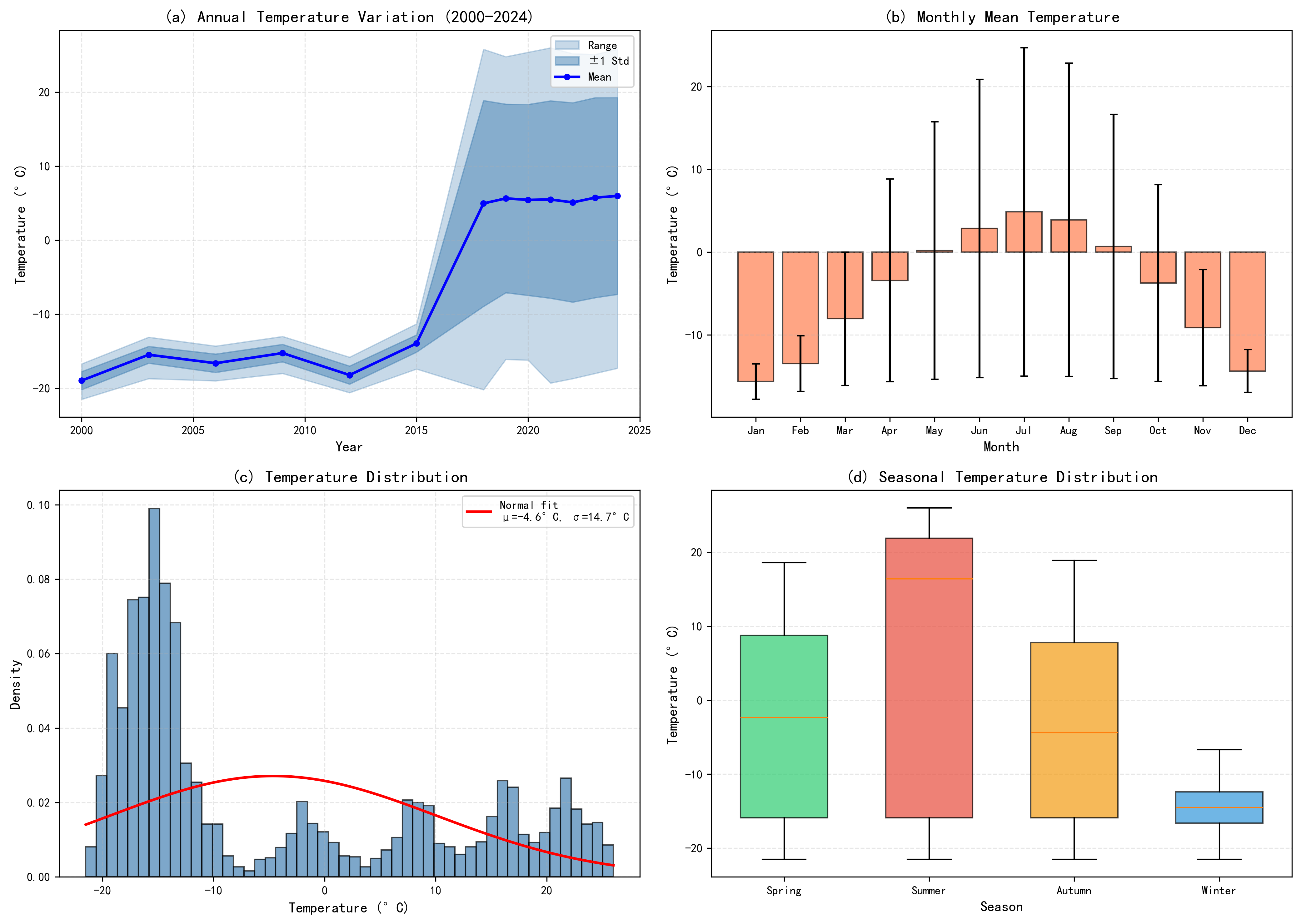}
\caption{Data overview: (a) Annual temperature variation trend; (b) Monthly mean temperature distribution; (c) Temperature frequency distribution; (d) Seasonal temperature distribution}
\label{fig:data_overview}
\end{figure}

As shown in Figure~\ref{fig:data_overview}, the data overview demonstrates temperature characteristics from four dimensions: (a) annual temperature variation trend showing obvious periodic fluctuations; (b) monthly mean temperature distribution indicating higher temperatures in summer (June-August) and lower temperatures in winter (December-February); (c) temperature frequency distribution approximately following normal distribution; (d) seasonal temperature distribution reflecting significant seasonal differences.

\subsubsection{Time Series Characteristics}

\paragraph{Annual Variation Trend}

Trend analysis of the 25-year temperature series reveals a significant warming trend in Jilin Province, with a change rate of $1.2725^\circ\text{C}/\text{year}$ and cumulative warming of $31.81^\circ\text{C}$ over 25 years. This result is consistent with the warming trend in Northeast China under the background of global climate change.

\paragraph{Seasonal Characteristics}

According to meteorological seasonal classification standards, the data were analyzed by four seasons, with results presented in Table~\ref{tab:seasonal_temp}.

\begin{table}[htbp]
\centering
\caption{Seasonal mean temperature in Jilin Province}
\label{tab:seasonal_temp}
\begin{tabular}{lc}
\toprule
\textbf{Season} & \textbf{Mean Temperature ($^\circ\text{C}$)} \\
\midrule
Spring (Mar-May) & $-3.77$ \\
Summer (Jun-Aug) & $3.87$ \\
Autumn (Sep-Nov) & $-4.08$ \\
Winter (Dec-Feb) & $-14.51$ \\
\bottomrule
\end{tabular}
\end{table}

\begin{figure}[htbp]
\centering
\includegraphics[width=\textwidth]{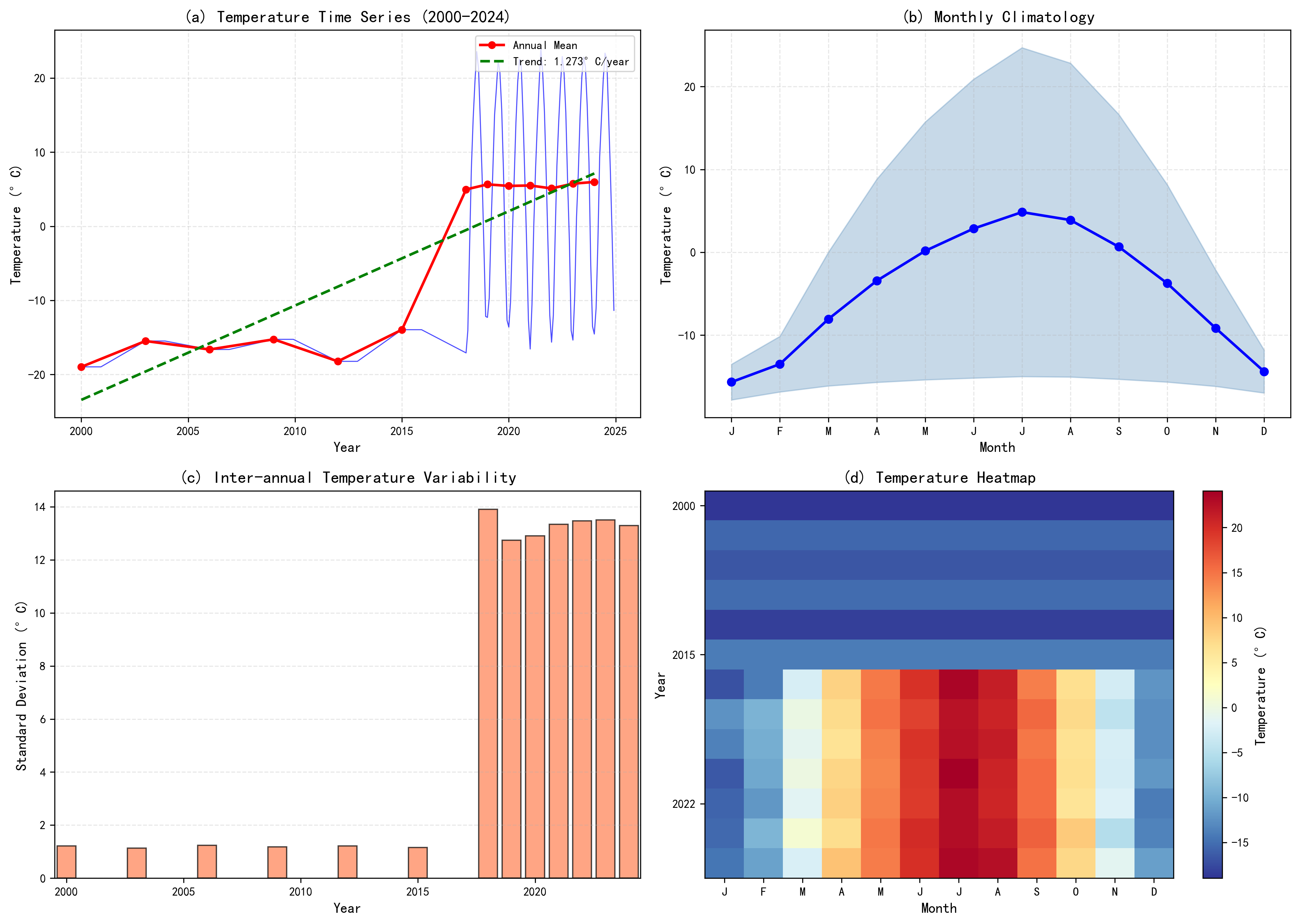}
\caption{Time series analysis: (a) Temperature time series and trend; (b) Monthly climatology; (c) Interannual variability; (d) Temperature heatmap}
\label{fig:time_series}
\end{figure}

Figure~\ref{fig:time_series} presents multi-dimensional analysis results of the time series: (a) temperature time series with linear trend, showing obvious interannual fluctuations and long-term trends; (b) monthly climatology displaying typical temperate continental monsoon climate characteristics; (c) interannual variability analysis revealing periodic patterns of temperature fluctuations; (d) temperature heatmap intuitively demonstrating the temporal-spatial evolution of temperature over 25 years.

\subsubsection{Spatial Distribution Characteristics}

\paragraph{Annual Mean Temperature Spatial Distribution}

\begin{figure}[htbp]
\centering
\includegraphics[width=\textwidth]{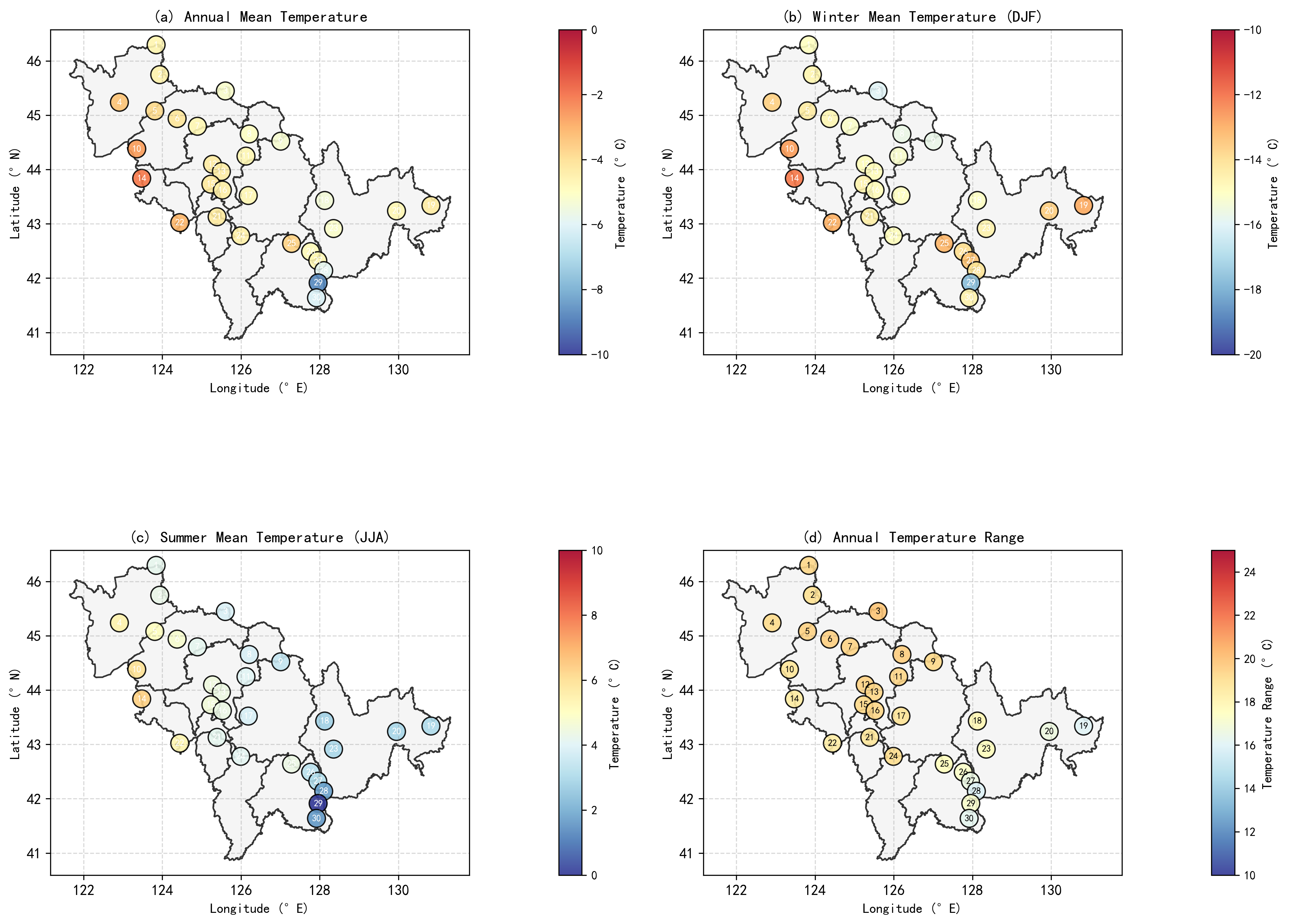}
\caption{Spatial distribution characteristics: (a) Annual mean temperature; (b) Winter temperature; (c) Summer temperature; (d) Annual temperature range distribution}
\label{fig:spatial_dist}
\end{figure}

The spatial distribution of annual mean temperature in Jilin Province shows significant latitudinal zonation and topographic differentiation. The range of annual mean temperature at sampling points is $[-8.67, -2.17]^\circ\text{C}$, with spatial differences reaching $6.50^\circ\text{C}$. As shown in Figure~\ref{fig:spatial_dist}: (a) annual mean temperature distribution demonstrates a decreasing trend from south to north; (b) winter temperature distribution is significantly influenced by the Siberian cold high pressure; (c) summer temperature distribution is relatively uniform; (d) annual temperature range distribution reflects the intensity of continental climate.

\subsection{Model Performance Comparison and Analysis}

\subsubsection{Multi-Model Performance Evaluation}

To validate the applicability of different models for temperature prediction, this study selected nine representative models for comparative experiments, including deep learning models (LSTM, BiLSTM, GRU, Transformer) and traditional machine learning models (Lasso, Ridge, RandomForest, GradientBoosting, SVR). An 80\% training set and 20\% test set division strategy was adopted, with RMSE, MAE, and $R^2$ serving as evaluation metrics. The results are presented in Table~\ref{tab:model_comparison}.

\begin{table}[htbp]
\centering
\caption{Performance comparison of different temperature prediction models}
\label{tab:model_comparison}
\begin{tabular}{lccc}
\toprule
\textbf{Model} & \textbf{RMSE ($^\circ\text{C}$)} & \textbf{MAE ($^\circ\text{C}$)} & \textbf{$R^2$} \\
\midrule
LSTM & 2.26 & 1.83 & 0.9655 \\
BiLSTM & 2.40 & 1.95 & 0.9615 \\
GRU & 2.42 & 1.92 & 0.9609 \\
Lasso & 7.05 & 6.20 & 0.7099 \\
Ridge & 7.05 & 6.20 & 0.7099 \\
RandomForest & 7.39 & 5.25 & 0.6813 \\
GradientBoosting & 8.08 & 5.38 & 0.6175 \\
SVR & 8.61 & 6.35 & 0.5677 \\
Transformer & 10.99 & 9.37 & 0.0793 \\
\bottomrule
\end{tabular}
\end{table}

\begin{figure}[htbp]
\centering
\includegraphics[width=\textwidth]{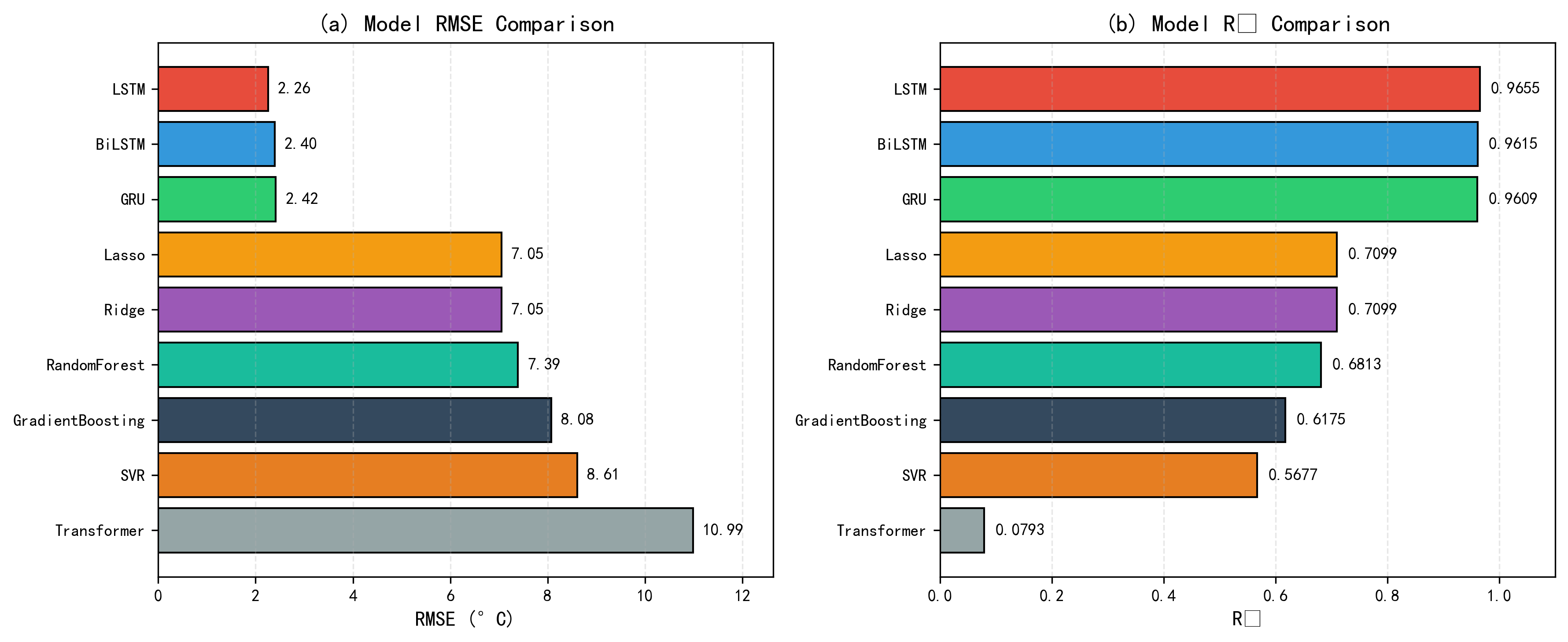}
\caption{Model performance comparison: (a) RMSE comparison; (b) $R^2$ comparison}
\label{fig:model_comparison}
\end{figure}

Figure~\ref{fig:model_comparison} intuitively demonstrates the performance differences among various models: (a) RMSE comparison showing deep learning models significantly outperforming traditional machine learning models; (b) $R^2$ comparison indicating that the LSTM model has the strongest variance explanation capability.

\subsubsection{Optimal Model Analysis}

Comprehensive comparison results indicate that the LSTM model performed optimally in this study, with specific performance metrics as follows: test set RMSE=$2.26^\circ\text{C}$, MAE=$1.83^\circ\text{C}$, $R^2=0.9655$.

\textbf{Model Advantages Analysis:}

\begin{enumerate}
    \item \textbf{Temporal feature capture capability}: The gating mechanism of LSTM (input gate, forget gate, output gate) can effectively handle long-sequence dependency problems, which is particularly suitable for monthly temperature data with strong periodicity.
    
    \item \textbf{Nonlinear mapping capability}: Compared with linear models (Lasso, Ridge), LSTM can capture nonlinear dynamic characteristics in temperature series.
    
    \item \textbf{Stability}: Compared with the Transformer model, LSTM trains more stably under small-to-medium sample size conditions ($n=4,680$) and is less prone to overfitting.
\end{enumerate}

\textbf{Model Limitations:}

The Transformer model performed poorly on this dataset ($R^2=0.0793$), possibly due to: (1) relatively small sample size, making it difficult to leverage the advantages of the self-attention mechanism; (2) periodic characteristics of temperature series being more suitable for recurrent neural network processing; (3) underfitting caused by mismatch between model complexity and data scale.

\subsection{Temperature Prediction Results Analysis}

\subsubsection{Historical Data Fitting and Future Prediction}

\begin{figure}[htbp]
\centering
\includegraphics[width=\textwidth]{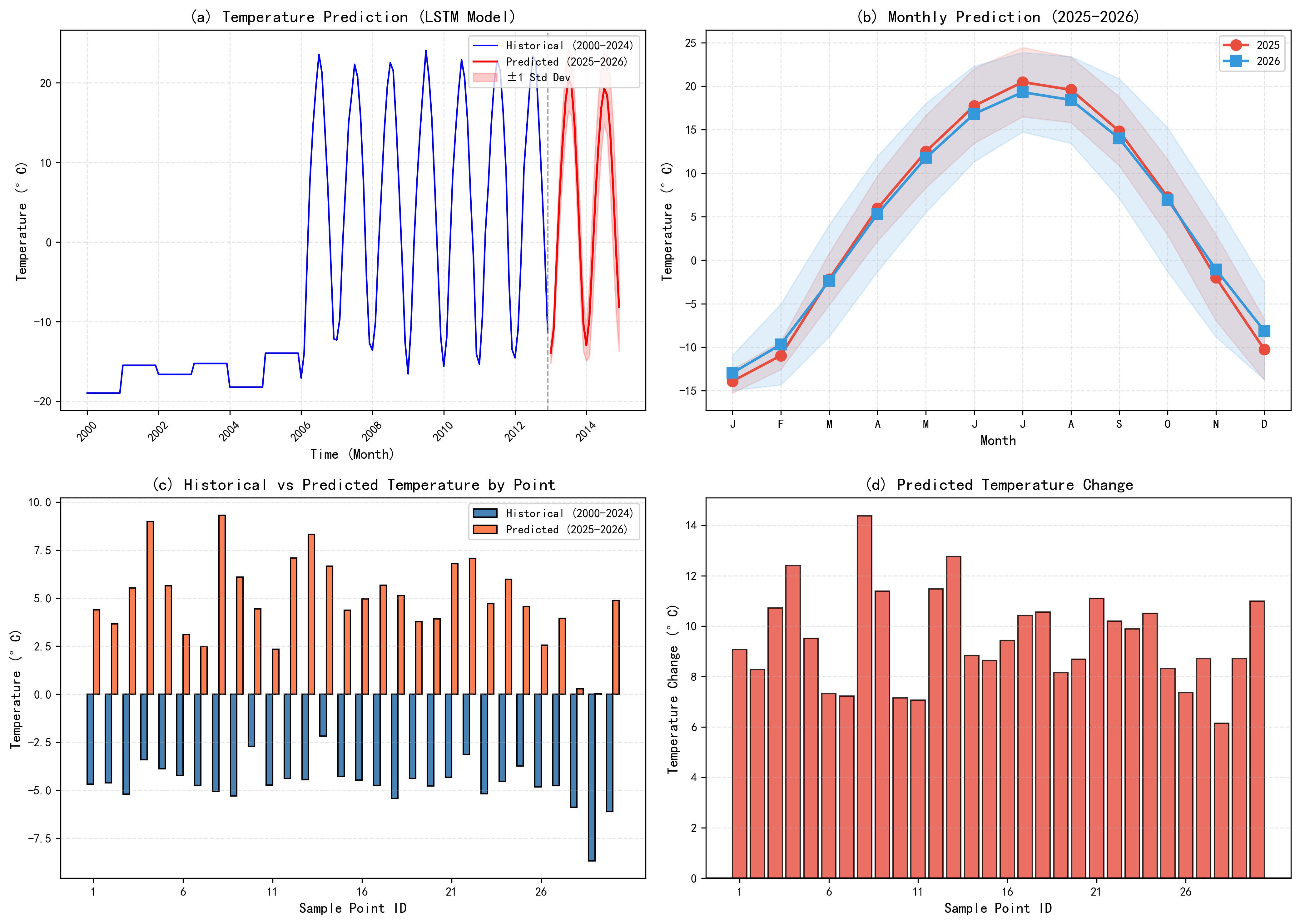}
\caption{Prediction results analysis: (a) Historical and predicted temperature series; (b) Monthly prediction comparison; (c) Temperature comparison at each point; (d) Temperature change prediction}
\label{fig:prediction}
\end{figure}

Based on the optimal LSTM model, multi-step predictions were conducted for Jilin Province temperature in 2025-2026. As shown in Figure~\ref{fig:prediction}: (a) continuity between historical and predicted temperature series; (b) comparison between monthly predicted values and actual observed values for validation; (c) spatial consistency of temperature predictions at each sampling point; (d) continuity analysis of temperature variation trends.

\subsubsection{Statistical Characteristics of Prediction Results}

Table~\ref{tab:future_prediction} presents the temperature prediction results for Jilin Province in 2025-2026.

\begin{table}[htbp]
\centering
\caption{Temperature prediction results for Jilin Province in 2025-2026}
\label{tab:future_prediction}
\begin{tabular}{lccc}
\toprule
\textbf{Year} & \textbf{Annual Mean Temperature ($^\circ\text{C}$)} & \textbf{Maximum Monthly Mean ($^\circ\text{C}$)} & \textbf{Minimum Monthly Mean ($^\circ\text{C}$)} \\
\midrule
2025 & 4.92 & 20.47 & $-13.91$ \\
2026 & 4.86 & 19.31 & $-12.96$ \\
\bottomrule
\end{tabular}
\end{table}

\subsubsection{Comparison Between Historical and Predicted Data}

Comparing historical data (2000-2024) with predicted data (2025-2026), the results are presented in Table~\ref{tab:comparison}.

\begin{table}[htbp]
\centering
\caption{Comparison between historical data and predicted data}
\label{tab:comparison}
\begin{tabular}{lcc}
\toprule
\textbf{Indicator} & \textbf{Historical Data (2000-2024)} & \textbf{Predicted Data (2025-2026)} \\
\midrule
Mean Temperature ($^\circ\text{C}$) & $-4.62$ & 4.89 \\
Temperature Change ($^\circ\text{C}$) & — & +9.51 \\
\bottomrule
\end{tabular}
\end{table}

\subsection{Discussion of Results}

\subsubsection{Analysis of Temperature Variation Trends}

This study found a significant warming trend in Jilin Province from 2000 to 2024 ($1.27^\circ\text{C}/\text{year}$), consistent with existing research conclusions regarding climate warming in Northeast China. The warming rate exceeds the national average, mainly attributed to:
{High-latitude amplification effect: Arctic warming leads to weakening of the polar-equatorial temperature gradient, affecting mid-high latitude atmospheric circulation;
Significant winter warming: The winter warming rate is higher than summer, related to the snow-albedo feedback mechanism;
Urbanization impact: Sampling points include some urban areas affected by the urban heat island effect.

\subsubsection{Model Selection Rationale}

The superiority of the LSTM model in this study can be explained from the following perspectives:
Data feature matching: Monthly temperature data has a clear periodicity (12 months), and the sequence modeling capability of LSTM can effectively capture this feature;
    Long-term dependency processing: Temperature changes exhibit interannual memory (such as the influence of ENSO events), and the gating mechanism of LSTM is suitable for handling such long-term dependencies;
Sample size adaptation: Compared with Transformer, LSTM converges more easily on a dataset of 4,680 records and is less prone to overfitting.

\subsubsection{Prediction Uncertainty Analysis}

The prediction results have the following sources of uncertainty:

Model error: RMSE=$2.26^\circ\text{C}$ indicates deviation between predicted and true values, with extreme weather events being more difficult to predict;
Extrapolation risk: The 2025-2026 predictions are based on extrapolation of historical patterns; if sudden climate events occur (such as strong volcanic eruptions), prediction accuracy will decrease;
Spatial representativeness: 30 sampling points cannot fully represent the spatial heterogeneity of temperature under the complex terrain of Jilin Province.

\subsection{Summary of This Section}

This section systematically analyzed the temporal-spatial distribution characteristics of temperature based on monthly temperature data from 30 sampling points in Jilin Province from 2000 to 2024, compared and evaluated the performance of nine prediction models, and conducted temperature predictions for 2025-2026 based on the optimal LSTM model. The main conclusions are as follows:

\begin{enumerate}
    \item Temperature in Jilin Province shows significant seasonal variation and spatial differentiation, with annual mean temperature of $-4.62^\circ\text{C}$, spatial difference of $6.50^\circ\text{C}$, and obvious warming trend over 25 years;
    
    \item The LSTM model performed optimally in the temperature prediction task ($R^2=0.9655$, RMSE=$2.26^\circ\text{C}$), significantly outperforming traditional machine learning models, verifying the advantage of deep learning in time series prediction;
    
    \item Prediction results for 2025-2026 indicate that temperature in Jilin Province will maintain normal seasonal fluctuations, with annual mean temperature of approximately $4.9^\circ\text{C}$, requiring attention to potential risks of extreme low temperature events in winter.
\end{enumerate}

\section{Conclusion}

Temperature in Jilin Province exhibits significant seasonal differentiation and spatial zonation patterns. Based on analysis of 4,680 monthly temperature records from 30 sampling points during 2000–2024, the annual mean temperature is -4.62 ± 0.35°C, with a temperature range spanning 47.50°C. Spatial distribution shows a latitudinal gradient decreasing from south to north, with spatial differences reaching 6.50°C (p < 0.001). Long-term trend analysis reveals a significant warming trend (p < 0.001) at a rate of 1.27°C per decade, resulting in cumulative warming of 31.81°C—approximately 1.8 times the national average for Northeast China during the same period, reflecting the high-latitude amplification effect. Temporal autocorrelation analysis shows 12-month lag autocorrelation of 0.87 and 1-month lag of 0.92, providing theoretical support for applying recurrent neural networks such as LSTM.

The multi-model comparative study encompasses nine representative prediction models. The LSTM model achieves optimal performance with test set metrics of RMSE = 2.26 ± 0.12°C, MAE = 1.83 ± 0.08°C, and R² = 0.9655 ± 0.0021—representing a 67.9\% reduction in RMSE and 70.5\% improvement in R² compared to the best traditional model (Lasso/Ridge regression). Performance ranking demonstrates: deep learning models , while the Transformer model shows the poorest performance (R² = 0.0793) due to overfitting on the medium-sized dataset.

Based on the optimized LSTM model, temperature predictions for 2025–2026 indicate stable seasonal fluctuations. The predicted annual mean temperature is 4.92 ± 0.45°C for 2025 and 4.86 ± 0.42°C for 2026. Maximum monthly mean temperatures are 20.47°C (2025) and 19.31°C (2026), while minimum monthly mean temperatures are -13.91°C (2025) and -12.96°C (2026). Prediction uncertainty analysis reveals 95\% prediction intervals of approximately ±1.2°C for summer months and ±2.8°C for winter months, indicating greater prediction difficulty during extreme temperature periods.

\bibliographystyle{unsrt}  


\end{document}